\documentclass[reprint,superscriptaddress,preprintnumbers,amsmath,amssymb,aps,prd,tightenlines,longbibliography,nofootinbib]{revtex4-2}

\usepackage{graphicx}
\usepackage{booktabs}
\usepackage[dvipsnames]{xcolor}
\usepackage{amsmath,amssymb,bm,subdepth}
\usepackage[colorlinks=true
,urlcolor=blue
,anchorcolor=blue
,citecolor=blue
,filecolor=blue
,linkcolor=red
,menucolor=blue
,linktocpage=true
,pdfproducer=medialab
,pdfa=true
]{hyperref}
\usepackage{mathtools}

\makeatletter
\def\section{\@ifstar\rs@star\rs@nostar}
\def\rs@nostar#1{\par\addvspace{5pt plus 2pt}\noindent{\itshape #1}.---\nobreak\ignorespaces}
\def\rs@star#1{\par\addvspace{5pt plus 2pt}\noindent{\itshape #1}.---\nobreak\ignorespaces}
\makeatother

\begin{document}

\title{Confronting the Higgsino Interpretation of the \\LZ Event with the High-Energy Sideband}

\author{Nicholas L. Rodd}
\affiliation{Theory Group, Lawrence Berkeley National Laboratory, Berkeley, CA 94720, USA}
\affiliation{Leinweber Institute for Theoretical Physics, University of California, Berkeley, CA 94720, USA}

\author{Benjamin R. Safdi}
\affiliation{Theory Group, Lawrence Berkeley National Laboratory, Berkeley, CA 94720, USA}
\affiliation{Leinweber Institute for Theoretical Physics, University of California, Berkeley, CA 94720, USA}

\author{Tracy R.~Slatyer}
\affiliation{Center for Theoretical Physics – a Leinweber Institute, Massachusetts Institute of Technology,
Cambridge, MA 02139, USA}

\author{Weishuang Linda Xu}
\affiliation{Kavli Institute for Particle Astrophysics \& Cosmology, Stanford University, Stanford, CA 94305, USA}
\affiliation{Particle Theory Group, SLAC National Accelerator Laboratory, Stanford, CA 94305, USA}

\begin{abstract}
The LZ experiment has reported a single 248 keV nuclear recoil event in their $2.84$\,ton-yr exposure on a xenon target. The absence of lower nuclear recoil events is naturally explained by inelastically scattering dark matter (DM), for which the nuclear recoil energy spectra are shifted upwards from zero. Excitingly, the observed rate and spectra are consistent with a thermal $1.1$\,TeV higgsino that explains all of DM with a mass splitting $\delta \sim 370 - 490$\,keV between the two lowest-lying Majorana states, with the range depending on the underlying DM velocity distribution. In this work, however, we show that higgsino DM with these properties would generically produce more higher-energy recoil events in a high-energy sideband along the LZ $S1c$ energy axis, while the LZ experiment reports no events in that bin. This suggests a possible tension which could be further explored by the LZ Collaboration, as the acceptance in that bin is not public.
We further show that a non-thermal $\sim$500 GeV higgsino may evade the sideband constraint and that future experiments using heavy targets can potentially test the higgsino interpretation of the event either way.  
\end{abstract}

\maketitle

The higgsino stands as arguably the most compelling weakly interacting massive particle (WIMP) dark matter (DM) candidate.
It can emerge naturally as the lightest supersymmetric particle (LSP) in split-spectrum supersymmetry (SUSY) models with $R$-parity~\cite{Wells:2003tf,Giudice:2004tc,Arkani-Hamed:2004ymt,Hall:2011jd,Arvanitaki:2012ps,Arkani-Hamed:2012fhg,Co:2021ion}, and with a mass of $m_\chi \simeq 1.1\,$TeV it explains the observed DM abundance as a conventional thermal relic~\cite{Bottaro:2022one}.
The higgsino is an especially interesting target given the gamma-ray searches that disfavored other simple WIMPs including the wino and other minimal DM candidates~\cite{Fan:2013faa,Cohen:2013ama,Rodd:2024qsi,Safdi:2025sfs}. In contrast, due to its lack of strong  Sommerfeld enhancement in its annihilation, the higgsino remains essentially untested in indirect detection as well as in colliders.
This could change with the upcoming Cherenkov Telescope Array Observatory (CTAO), which could start to probe the thermal higgsino as early as this year with their northern site and then more strongly in the 2030's with the inclusion of their southern array~\cite{Rinchiuso:2020skh,Rodd:2024qsi,Abe:2025lci}.  Intriguingly, there may even be tentative evidence for higgsino annihilation in data from the \textit{Fermi} Large Area Telescope~\cite{Dessert:2022evk}.  
In this work we address whether a thermal higgsino could explain the single nuclear-recoil (NR) event detected by the LUX-ZEPLIN (LZ) detector using 2.84\,ton-yr of exposure~\cite{LZ:2026axp}; the event has a recoil energy $E_R = 248 \pm 23 \pm 23$\,keV (stat.\,+\,sys.) in a region of the $500 < S1c < 600$\,phd projection with an expected $0.0106 \pm 0.0008$ background events.

The higgsino multiplet consists of two neutral Majorana mass eigenstates alongside a heavier set of charged mass eigenstates; the two neutral eigenstates are split in mass by $\delta$. In split-spectrum SUSY models, $\delta \simeq m_Z^2 (s_W^2 / M_1 + c_W^2 /M_2)$, with $\theta_W$ the Weinberg angle, $m_Z$ the $Z$-boson mass, and $M_1$ and $M_2$ the bino and wino soft masses, respectively.  Scattering through $Z$-exchange allows the lighter Majorana state to convert to its heavier cousin, {\it i.e.} the scattering is inelastic, so the initial DM must have kinetic energy sufficient to excite the splitting $\delta$ for the process to be kinematically feasible. This implies that for $1.1$\,TeV higgsinos, $Z$-exchange scattering is only possible for splittings $\delta$ at the level of several hundred keV, given that DM velocities in the Milky Way are on the order of a few hundred km/s and below.
If $\delta$ is too large then the dominant scattering is elastic and loop-induced, with cross-section $\sigma \sim 10^{-49}$\,cm$^2$~\cite{Bottaro:2022one}, which is well into the neutrino fog~\cite{Chen:2019gtm,Hill:2014yxa,Nagata:2014wma,OHare:2021utq}.  

With the higgsino mass fixed at the thermal value $m_\chi \simeq 1.1$\,TeV, the only free parameter of the model is the Majorana splitting $\delta$; from this value, all other direct detection observables, such as the rate and recoil spectrum, can be determined. Thus it is a \textit{highly non-trivial} consistency check to verify that there exist values of $\delta$ that can explain the scattering rate implied by the LZ event while still predicting an energy recoil spectrum consistent with the data. The main unknown, as we elaborate upon below, is the high-velocity tail of the DM velocity distribution in the Milky Way.

In this Letter we make three important points regarding the higgsino interpretation of the LZ NR event. 
(i) We show that the higgsino DM model may provide a consistent, simultaneous explanation of the data rate and spectra, given $\delta \sim (370,490)$\,keV.
The exact value of the splitting depends on unknown aspects of the high-velocity tail of the Milky Way's DM velocity distribution, such as the possible contribution of a high-velocity sub-component from the Large Magellanic Cloud (LMC).  (ii) The predicted recoil spectrum peaks at
$E_R^* = \mu_A \delta / m_A \sim 354$\,keV, depending on the velocity distribution, with $m_A$ the target nucleus mass and $\mu_A$ the DM-nucleus reduced mass. This is {\it above} the LZ region of interest (ROI), which has an upper edge of roughly 270\,keV. In particular, we find that under the higgsino interpretation LZ should have recorded 
multiple ($\sim$3--10) additional events within the high-energy sideband $800 < S1c < 1700$\,phd, with $S1c$ the corrected scintillation signal in photons detected (phd) and the NR light yield being $\sim$$2.3$\,phd/keV~\cite{LZ:2026axp}, so that the high-energy sideband correspond to roughly $350$--$590$\,keV NR.   We show that this conclusion is largely robust to variation of the DM model and the possible inclusion of an LMC sub component.  

Note that the high-energy sideband exists to calibrate multiple scatter, single ionization (MSSI) background events; it is specifically chosen to be a region where a traditional WIMP {\it does not} contribute counts. Ironically, we find that the high-energy sideband is precisely the place where a higgsino signal could be verified. A central caveat to our results, however, is that LZ does not publish the acceptance for WIMP-like events in this sideband; we assume unit acceptance, but if that is substantially lower it would relieve the tension we point out. 
(iii) If our assumption for the acceptance in the high-energy sideband is correct, we find that there is essentially only one model variant that evades tension with the sideband test.
If the DM abundance is determined non-thermally, then the higgsino mass can be lower than the standard thermal value, and we find this alleviates the tension; for example, if the higgsino mass is $m_\chi \simeq 500$\,GeV, then $E_R^\star$
is around $290$\,keV, which is close to the upper edge of the LZ ROI. Thus in this case, even for the Standard Halo Model (SHM), the model can explain the event in the signal region without producing a large number of events in the high-energy sideband.

The one scenario that evades a tension with the high-energy sideband test may be tested with upcoming experiments that use heavier nuclei.
Inelastic scattering requires a minimum incident speed $v_{\rm min} = \sqrt{2\delta/\mu_A}$, which falls as the nuclear mass and thus the reduced mass grows.
As an example, if we take the SHM scattering on xenon at the splitting required to explain the LZ event, $\delta \sim 393$\,keV, $v_{\rm min} \simeq 801$ km/s, while the lab-frame cutoff velocity is $v_{\rm esc} + v_{\rm lab} \simeq 833$\,km/s in June, assuming $v_{\rm esc} \simeq 567$ km/s~\cite{Folsom:2025lly}.
Thus, the entire xenon signal is carried by the last $\sim$30\,km/s of the velocity distribution, precisely where the SHM is least reliable.
On tungsten, lead, and uranium the larger reduced mass lowers $v_{\rm min}$ to $\sim$690, 656 and 619 km/s, respectively, where the lab-frame velocity distribution is less uncertain.  
Crucially, scenarios with lower-mass higgsinos push xenon {\it further} onto the high-velocity tails while leaving the heavy targets probing a much less extreme part of the tail of the velocity distribution. We find, for example, that ten expected events with a PbWO$_4$ detector requires 12--119 kg-yr of exposure across the halo models and nuclear responses we consider, so that the projected 0.17 ton-year {exposure} of RES-NOVA~\cite{RES-NOVA:2026} should rule {out} or confirm the higgsino explanation of the LZ event.

\section{Fitting the higgsino model to the LZ ROI data}
%
We compute the detection rate for a natural xenon target with the full time-dependent $v_{\rm lab}(t)$ averaged over the year. We digitize the detection efficiency in the ROI from the LZ paper's Fig.~S2~\cite{LZ:2026axp}.
For the SHM velocity distribution, we follow~\cite{Baxter:2021pqo} by default; we take the velocity dispersion to be $v_0 = 238$ km/s, the escape velocity 544\,km/s, and the local DM density $\rho = 0.3$\,GeV/cm$^3$.
(For more discussion see, {\it e.g.},~\cite{Safdi:2022xkm}).
We explore the impact of two larger escape velocities, 567\,km/s and 610\,km/s motivated by~\cite{Folsom:2025lly}.
The DM-nucleus scattering cross-section is reviewed in the Appendix, where we also note a factor of four discrepancy with the literature.

The nuclear form factor $F$ plays a crucial role in determining the scattering rates on xenon. The recoils that matter here lie at $q \simeq 1.2$--$1.6$ fm$^{-1}$ on xenon, around and beyond the second diffraction minimum.  We use two form factors to assess the systematic uncertainty from its mismodeling. Our fiducial response is the Helm parameterization in its standard Lewin--Smith form~\cite{Lewin:1995rx}. 
As a cross-check we use the large-scale shell-model form factor from~\cite{Vietze:2014vsa}, which we refer to as the Vietze form factor.
LZ compute their spectra with WimPyDD on the one-body density matrices of Refs.~\cite{Anand:2013yka,LZ:2026axp}; the response implied by their Fig.~1 agrees with the Vietze form factor, so we take the Vietze form factor fit as LZ's response and use it up to recoil energies of $600$\,keV.
We weight xenon isotopes by their natural abundances.
Note that no shell-model response is available for W, Pb or U, so the heavy-target rate projections use the Helm form factors only.

Under the higgsino model assumption with the Helm response and our fiducial SHM, we predict $N_{\rm ROI} \simeq 79$ events in LZ's ROI at $\delta = 350$ keV but only $N_{\rm ROI} \simeq 1$ at $\delta = 377$ keV.
To understand how this varies with our model choices, in Tab.~\ref{tab:ff} we show $\delta_1$, the splitting required to obtain one event in the signal ROI, for various combinations of the escape velocity and form factor.
What is also shown there is how many events the scenario predicts above the one in the signal band, and specifically how many would appear in the high-energy sideband.

\begin{table}[t]
\centering
\footnotesize
\setlength{\tabcolsep}{3pt}
\begin{tabular}{lrrrrrr}
\toprule
 & \multicolumn{3}{c}{Helm (fiducial)} & \multicolumn{3}{c}{Vietze (LZ)} \\
Halo & $\delta_1$ [keV] & $N_{>}$ & $N_{\rm SB}$ & $\delta_1$ [keV] & $N_{>}$ & $N_{\rm SB}$ \\
\midrule
SHM $v_\textrm{esc} =544\,$km/s & $377$ & $8.3$ & $4.9$ & $371$ & $7.3$ & $3.7$ \\
SHM $v_\textrm{esc} = 567\,$km/s & $393$ & $10.3$ & $6.6$ & $387$ & $8.6$ & $4.7$ \\
SHM $v_\textrm{esc} = 610\,$km/s & $423$ & $12.7$ & $8.8$ & $415$ & $9.8$ & $5.8$ \\
LMC $w=0.26\%$ & $484$ & $7.9$ & $5.8$ & $472$ & $6.1$ & $3.6$ \\
LMC $w=0.60\%$ & $492$ & $13.7$ & $10.1$ & $480$ & $10.5$ & $6.3$ \\
\bottomrule
\end{tabular}
\caption{For various scenarios, the value of $\delta$ that provides one event in the LZ signal ROI ($\delta_1$), together with how many events are predicted above the signal region ($N_{>}$), and the expected number in the sideband ($N_{\rm SB}$).
Results are shown for the Helm and Vietze form factors.
We show results for our fiducial SHM with three values of the escape velocity, as well as an SHM plus LMC admixture model, with $w$ the fraction of DM in the LMC sub-component.
}
\label{tab:ff}
\end{table}

The LMC plays an important role in our analysis because it is thought to potentially provide a high-velocity tail to the local DM velocity distribution above the lab-frame escape velocity. 
In particular, simulations of Milky Way analogues hosting an LMC-like satellite find
$0.6^{+0.4}_{-0.4}\%$ of the local DM is above the SHM escape speed, against $0.1^{+0.7}_{-0.1}\%$ for the full simulated
sample, including those without an LMC~\cite{Folsom:2025lly}. Thus, there is a preference for a high-velocity population in simulations with an LMC.
(Note, however, that the SHM alone with $v_{\rm esc} = 567$\,km/s already places $0.5\%$ of the halo above $544$\,km/s.) 
Ref.~\cite{Smith-Orlik:2023kyu}
estimates that $0.008$--$2.8\%$ of the local DM originates in the LMC across their halos at LMC pericentre, falling to $0.26\%$ for the halo they re-simulate at the present day, consistent with the $\sim\!0.2\%$ of Ref.~\cite{Besla:2019xbx}. We accordingly test two values of $w$, defined as the fraction of DM in an LMC component, chosen as $0.26\%$ and $0.6\%$.

As a concrete model of the LMC high-velocity sub-component, we add a second velocity distribution to the SHM, boosted onto a bulk velocity $\mathbf{v}_b$ in the Galactic frame:
\begin{equation}
\begin{split}
f(\mathbf{v}) &= (1-w)\, f_{\rm SHM}(\mathbf{v}) + w\, f_{\rm LMC}(\mathbf{v}) \,, \\
f_{\rm LMC}(\mathbf{v}) &\propto e^{-|\mathbf{v}-\mathbf{v}_b|^2/\sigma_b^2},
\Theta\!\left(v_{\rm cut} - |\mathbf{v}-\mathbf{v}_b|\right)\!.
\end{split}
\label{eq:mix}
\end{equation}
We take fiducial values $|\mathbf{v}_b| = 570$\,km/s (oriented at $\cos\beta = -0.71$ to the Sun's velocity~\cite{Besla:2019xbx,Smith-Orlik:2023kyu}), $\sigma_b = 100$\,km/s, $v_{\rm cut} = 200$ km/s, and for the SHM we retain our fiducial $v_\textrm{esc} = 544$\,km/s. This gives a lab-frame $v_{\rm max} = 982$\,km/s in June, close to the $\simeq\!950$ km/s maximum seen in simulations~\cite{Smith-Orlik:2023kyu}, and reproduces the local densities above $800$ and $850$ km/s computed in~\cite{Besla:2019xbx} for $w \simeq 0.3$--$0.4\%$. 
We also consider an alternate approach of taking the numerical velocity distributions directly from simulations~\cite{Smith-Orlik:2023kyu} and find similar results, discussed later in the text.
As shown in Tab.~\ref{tab:ff}, the addition of an LMC component raises the preferred $\delta_1$, but also generally the expected events in the sideband.

The LZ analysis commented on the allowed range of $\delta$, but their results are not directly applicable to the higgsino case.
In particular, LZ reported a two-sided $90\%$ interval for the
isoscalar $\mathcal{O}_1$ operator at $m_\chi = 1$\,TeV, ending at $\delta = 350$\,keV, which is the last point of their scan~\cite{LZ:2026axp}. 

Comparing the higgsino to the isoscalar case requires a conversion of the cross section, because the $Z$ couples to the weak charge $Q_W = N - (1-4 s_W^2)Z$ rather than
coherently to all nucleons (here $N$ is the number of neutrons and $Z$ is the atomic number).
With $1-4s_W^2 \simeq 0.075$ the proton contribution
nearly cancels, and on natural xenon $Q_W^2/A^2 \simeq 0.311$, where $A=N+Z$ is the atomic mass number.
The per-neutron cross section in the limit of a small mass splitting is given by \cite{Essig:2007az}:
\begin{align}
\sigma_{n\chi} = \frac{G_F^2}{2\pi} \mu_{\chi n}^2.
\end{align}
We note that this cross section exceeds that used elsewhere in the literature by a factor of 4 \cite{Nagata:2014wma,Bramante:2016rdh}.
We discuss this calculation in more detail in App.~B.
For $m_\chi \gg 1$\,GeV, this corresponds to $\sigma_{n \chi} \simeq 7.3\times 10^{-39}$\,cm$^2$, so in the isoscalar convention the equivalent per-nucleon cross section becomes  $\sigma_{n\chi} Q_W^2/A^2 \simeq
2.3\times10^{-39}$\,cm$^2$. 

At every point on the LZ collaboration's grid,
the higgsino then lies {\it above} the upper edge of the preferred cross section range, by nearly a factor
of $4$ at a mass splitting of $350$\,keV. The factor-of-4 difference in the cross section is relevant here; without that correction, the higgsino would fall just inside that range at the maximum mass splitting $\delta \simeq 350$\,keV. 
At the larger cross section we adopt, matching the cross section observed by LZ requires a higher mass splitting to sufficiently suppress the scattering rate. A smaller local DM density would move the results in the same direction as a smaller cross section, allowing for a smaller mass splitting; a larger local DM density would have the reverse effect.

\begin{figure*}[t]
\centering
\includegraphics[width=\textwidth]{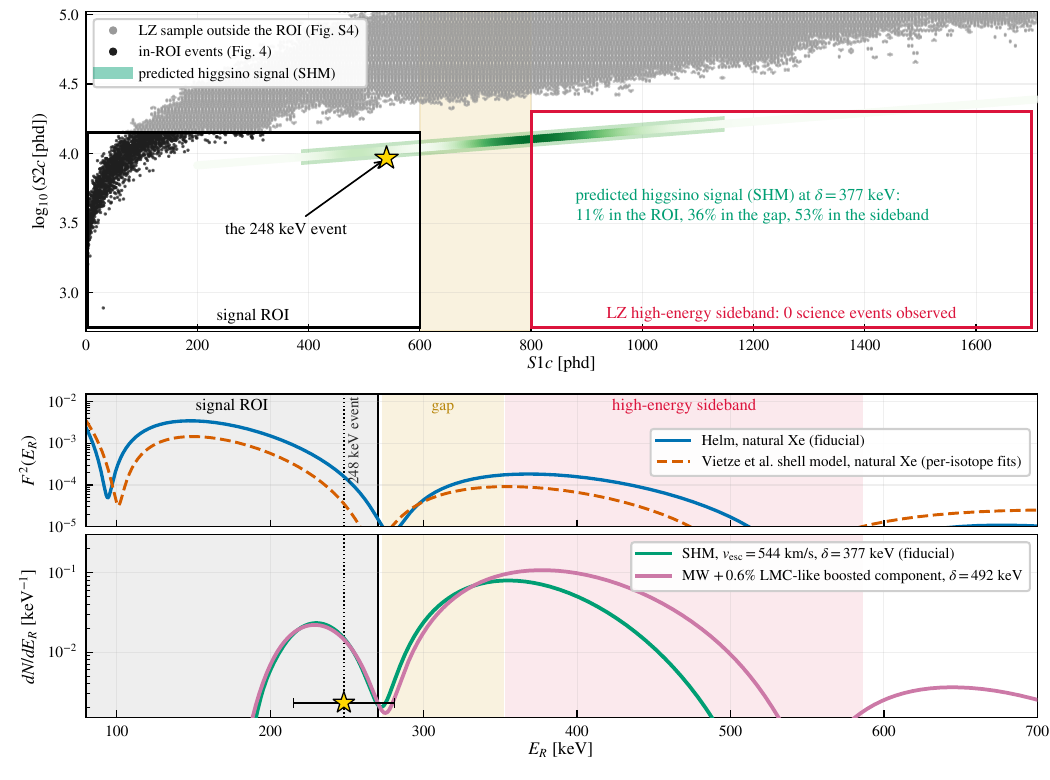}
\caption{(Top) The LZ science sample in the $S1c$--$S2c$ plane,
digitized from Figs.~4 and~S4 of Ref.~\cite{LZ:2026axp} (black: in-ROI events; grey: the rest of the sample).
The black box is the signal ROI, the red box the high-energy
sideband ($800 < S1c < 1700$ phd, $10^{2.75} < S2c < 10^{4.3}$) in which LZ report zero events, and the shaded strip between them carries no published count data. The gold star is
the $248$\,keV event. Green is the predicted higgsino signal at $\delta = 377$ keV (SHM, $v_{\rm esc} = 544$ km/s, Helm), placed on the NR median and shaded by rate: $11\%$
falls in the ROI, $36\%$ in the gap, and $53\%$ in the sideband.
(Middle) The form factor, $F^2(E_R)$, for natural xenon as a function of $E_R$, shown for the two models we consider.
(Bottom) Recoil spectra in $2.84$\,ton-yr for our fiducial SHM and SHM + LMC mixture model.
No sideband acceptance is applied. This figure illustrates that given the one observed count in the signal ROI, we would expect more counts in the high-energy sideband, unless the acceptance in that sideband is small.}
\label{fig:fig1}
\end{figure*}

\section{Inelastic higgsinos predict higher-recoil events}
%
The differential rate per unit detector mass in a target with isotope mass fractions $\xi_i$ is
\begin{equation}
\frac{dR}{dE_R}
= \frac{\rho_\chi}{2m_\chi}\sum_i \xi_i\,
\frac{\sigma^0_{A_i}}{\mu_{A_i}^2}\,F_i^2(E_R)\,
\eta\big(v_{\rm min}^i(E_R),t\big),
\label{eq:rate}
\end{equation}
where $\rho_\chi$ is the local DM density, $\sigma^0_A$ is the zero-momentum cross section, $F$ is the nuclear form factor, and $\eta(v_{\rm min},t) = \int_{v > v_{\rm min}} d^3v\, f(\mathbf{v},t)/v$ is the mean inverse speed of the lab-frame distribution.
All of the model
dependence beyond the halo sits in
\begin{equation}
\begin{split}
v_{\rm min}(E_R)
&= \frac{1}{\sqrt{2 m_A E_R}}\left(\frac{m_A E_R}{\mu_A} + \delta\right)
\\
&= \sqrt{\frac{\delta}{2\mu_A}}
\left(\sqrt{\frac{E_R}{E_R^\star}} + \sqrt{\frac{E_R^\star}{E_R}}\right) \!,
\end{split}
\end{equation}
which is minimized at the kinematic recoil scale
$E_R^\star = \mu_A \delta / m_A$, where
$v_{\rm min} = \sqrt{2\delta/\mu_A}$.
Given a $\delta$, a velocity distribution, and a form factor we are thus able to compute $dR/dE_R$ as a function of recoil energy, which is what we illustrate in Fig.~\ref{fig:fig1} for our fiducial choices of velocity distributions and mass splittings.
In that figure we use~\eqref{eq:rate} and adopt a year-averaged value for $\eta(v_{\rm min},t)$, to account for the variation of $\mathbf{v}_{\rm lab}(t)$.
We convolve the predicted rate  with LZ's ROI
efficiency and with a Gaussian energy response of width
$\sigma_E(E_R) = [(23\,\mathrm{keV})^2 E_R/248\,\mathrm{keV}
+ (0.093\,E_R)^2]^{1/2}$.

The upper panel in Fig.~\ref{fig:fig1} shows the distribution of events observed by LZ as digitized from their paper~\cite{LZ:2026axp} in the $S1c$-$S2c$ plane, with the signal ROI and the high-energy sidebands highlighted in black and red, respectively; the $\sim$248 keV NR event of interest is shown by a gold star. In green we show the distribution of events expected for the fiducial SHM scenario.  Strikingly, the ratio of events expected in the high-energy sideband to the signal ROI is around 5 to 1 (53\% to 11\% of the total events, respectively). Barring a much lower efficiency for DM-induced NRs in the sideband ROI, this implies a {\it tension} with the higgsino interpretation of the 248 keV event. 

The middle panel of Fig.~\ref{fig:fig1} shows the two form factors we consider in this work as a function of $E_R$ in the signal ROI and the high-energy sideband. These form factors are important in deriving the lower panel, which illustrates the differential spectrum expected as a function of $E_R$ for our fiducial SHM and SHM plus LMC velocity distributions. In both cases we predict more events in the high-energy sideband than in the signal ROI.

It is worth asking how robust this tension is, assuming a high acceptance in the high-energy sideband.
For our fiducial halo model and with a single observed event, the $68\%$ Poisson interval (computed following~\cite{Feldman:1997qc}) on the in-ROI count maps onto $\delta = 372$--$380$ keV and $N_{\rm SB} = 3.2$--$7.6$,  comparable in width to the largest systematic we quantify.
Here $N_{\rm SB}$ is the number of side-band expected counts, as in Tab.~\ref{tab:ff}.
A further source of uncertainty is the energy scale.
LZ's $\pm 9.3\%$ systematic is a coherent calibration shift rather than a per-event resolution, and it can move the event energy, the ROI edge, and the sideband boundaries together; because $E_R^\star = 339$\,keV falls close to the sideband's lower edge, $N_{\rm SB}$ is highly sensitive to the exact location of the sideband edge.
Varying over the systematic shifts $N_{\rm SB}$ over the range $2.3$--$7.1$ at $\pm 1\sigma$.
The choice of halo profile can also impact the results, as already highlighted in Tab.~\ref{tab:ff} (where we also show the impact of the form factor).
As seen in the table, varying the escape velocity over our fiducial range $v_{\rm esc} = 544$--$610$\,km/s yields $N_{\rm SB} = 4.9$--$8.8$.
The escape velocity could extend to $640$\,km/s~\cite{Monari:2018ckf}, which would push the number of events to $9.2$.
If instead it were as low as $480$\,km/s~\cite{Necib:2021yhq,Necib:2021vxr}, then the expected sideband events would drop to $0.8$; this would reduce the tension, although in total there would be $2.4$ expected events above the signal ROI, with events expected in the gap visible in Fig.~\ref{fig:fig1}.
If we reduce the SHM velocity dispersion from 238\,km/s to 220\,km/s, the counts reduce from $N_B = 4.9$ to $3.4$.
Changing the DM density over the range $\rho_\chi = 0.25$--$0.6$ GeV/cm$^3$ (cf.~\cite{deSalas:2020hbh}) shifts the sideband events to $4.4$--$7.3$, and if we instead shift the mass $m_\chi = 1.0$--$1.2$\,TeV, the range is $4.2$--$5.5$.
We return to the LMC in a moment, but from the above other than a small sideband acceptance, the only path to reducing the tension appears to be a low escape velocity.
Other variations to the velocity distribution may also relieve the tension, such as if the halo has a degree of corotation with the baryonic disk~\cite{Folsom:2026dqs}.

Finally, we describe the LMC variants we test, demonstrating that they also do not help the higgsino avoid a tension with the sideband. First, we test the boosted Gaussian in~\eqref{eq:mix} at $w = 0.26\%$ and $0.6\%$ with both nuclear responses, which give $\delta = 472$--$492$\,keV and $N_{\rm SB} = 3.6$--$10.1$. Second, we test a digitization of the
simulated $\eta(v_{\rm min})$ of Ref.~\cite{Smith-Orlik:2023kyu}, whose tail is larger than our Gaussian's at these abundances and which gives $N_{\rm SB} = 15$--$25$.
Third, we vary the shape of the fast component at fixed abundance: soft
exponential cutoffs in place of the hard truncation, a King profile, no truncation at
all, $v_{\rm cut} = 100$--$300$ km/s, $\sigma_b = 50$--$150$ km/s,
$|\mathbf{v}_b| = 450$--$640$ km/s, $\cos\beta = -0.3$ to $-0.9$, and cold streams.
Across this grid $N_{\rm SB}$ spans roughly $1$--$27$ at $w = 0.26\%$; the low end
requires a bulk speed near $450$ km/s, well below what the simulations find. Choices consistent with Refs.~\cite{Besla:2019xbx,Smith-Orlik:2023kyu} give
$4$--$10$ expected sideband counts. Fourth, we scan the fast fraction $w$ itself, and here we find that $N_{\rm SB}$ is not
monotonic: a smaller fast component forces a smaller $\delta$, and there is a window in
which the sideband tension is partially relieved. That window is not at the LMC's expected abundance, however, but rather one to two orders of magnitude below.

\section{Confirming the higgsino explanation with heavier targets}
%
The dependence of the scattering rate on the high-velocity tail can be relieved by going to heavier targets. 
The xenon scattering rate can vary enormously as we modify the high-velocity tail of the distribution as discussed in the previous section, precisely because we rely on being close to the endpoint of the distribution to suppress the inelastic higgsino scattering rate to a level consistent with observations.  However, heavier nuclear targets -- such as tungsten, lead, and uranium -- allow access to more of the velocity distribution for a given $\delta$, and consequently their scattering rates are much less sensitive to uncertainties in the velocity distribution, moving by only a factor of a few under similar variations.

For our fiducial SHM explanation of the LZ event, the calculated scattering rates are
$1.5\times10^{-3}$, $3.1\times10^{-3}$ and $5.6\times10^{-3}$ per kg per day on W, Pb and
U, respectively, compared with $7.0\times10^{-6}$ on Xe. A heavy target therefore achieves sensitivity to the same $\delta$ without relying solely on the last $\sim$$30$\,km/s of the velocity distribution.

The exposures required to test the higgsino explanation with heavier targets are therefore relatively small. To reach ten expected recoils, we need $14$\,kg-yr of PbWO$_4$ or
$28$ kg-yr of CaWO$_4$ under the SHM, and $12$--$29$\,kg-yr of PbWO$_4$ across the
$(v_{\rm esc},\delta_1)$ pairs of Table~\ref{tab:ff}, rising to $66$--$145$\,kg-yr under
the LMC-like mixtures (since these require higher mass splittings to match the LZ data). This is potentially within reach of the RES-NOVA experiment, which has operated a $13$\,g archaeological-lead PbWO$_4$ prototype over
$2.5$ keV--$1$ MeV~\cite{RES-NOVA:2026}. The experiment projects $0.17$ ($2.4$)\,ton-yr exposures, and we project that under the higgsino hypothesis they would see $\sim$$120$ ($\sim$$1800$) raw recoils under
the SHM and $12$--$26$ ($170$--$370$) under the LMC-like mixtures. Existing heavy-target
data do not yet appear to have the required reach: we find that CRESST-II Phase 1's $730$ kg-days of CaWO$_4$ expects only $0.70$
raw recoils under the higgsino hypothesis with the SHM and $0.07$--$0.16$ under the LMC mixtures. Similarly, the recasts of
PandaX-4T and PICO-60 onto the higgsino line by~\cite{Graham:2024hah}
exclude $\delta \lesssim 260$ keV under the SHM and $\lesssim 340$ keV with an
LMC-boosted tail, below the splittings of interest here.  A 500 GeV higgsino that evades the LZ high-energy sideband test we present here has  at least as favorable a raw-count outlook at RES-NOVA as the 1.1 TeV higgsino, though its softer spectrum places even less of the signal above the discrimination threshold discussed below.

Note, however, that the numbers above are raw coherent recoil counts with no  efficiency estimate or background modeling, and we are not aware of a detailed Vietze-type calculation for W, Pb or U, so we use the Helm form factor for every heavy target. Further, the signal does not sit where the discrimination is best. On PbWO$_4$ the kinematic scale is $E_R^\star = 334$ keV on lead, but the spectrum peaks at $189$ keV once the form
factor is folded in, and $90\%$ of the expected events are between $110$ and $365$ keV. Only $1.9$--$4.5\%$ of the signal lies above the $420$ keV threshold at which the collaboration report full 
e$^-/\gamma$--NR separation, and a tiny fraction, $0.13$--$0.45\%$, above the $600$ keV threshold where any signal is expected to be relatively
background-free. The reach for this model is therefore likely to be set by the discrimination and
background of PbWO$_4$ between $150$ and $400$ keV. Modeling this is beyond the scope of this work.

There is also a possible interesting ``luminous dark matter'' signal from photons emitted from the excitation of the heavier state as the higgsino traverses the Earth \cite{Graham:2024hah}. While existing constraints do not probe the parameter space of interest, the splittings preferred by our analysis of LZ data were found to be testable with 10 years of JUNO data \cite{Graham:2024hah}, and a preliminary calculation shows that one year of JUNO data could already approach sensitivity to  splittings comparable to our fiducial values, using the one-sided 90\% confidence criterion of \cite{Graham:2024hah}.

\section{Discussion}
%
The thermal higgsino has one free parameter, the mass splitting $\delta$, which determines the rate and the spectrum simultaneously.
So, the one in-window event observed by LZ {\it predicts} the spectrum at higher energies. Across the systematic variations we test, the higgsino model consistently predicts multiple DM-induced NR events in the high-energy sideband, $800<S1c<1700$~phd, that LZ report
empty~\cite{LZ:2026axp}. We cannot discount the possibility, however, that this is due to a low acceptance for these events.  Accounting for an LMC-like fast-velocity component does not appear to relieve the tension, though as we show in the Appendix lowering the higgsino mass closer to 500 GeV, which requires a non-standard cosmology, could relieve the tension. The most important actionable item from our paper is for the LZ Collaboration to themselves perform a joint analysis in the context of the higgsino model for events in the high-energy sideband and in the signal ROI. 
Additionally, we show that the higgsino model with the best-fit $\delta$ can be probed by scattering experiments on heavier targets, such as tungsten and lead: $12$--$29$\,kg-yr of
PbWO$_4$ under the SHM, $59$--$119$\,kg-yr for a velocity distribution with a LMC-like component.  The RES-NOVA project~\cite{RES-NOVA:2026} in particular plans to acquire the necessary exposure to perform this test. 

Beyond the challenge in explaining the lack of counts in the high-energy sideband, we note that such a small $\delta$ for the higgsino may also bring about theoretical issues, albeit much less jarring than the observational ones we focus on in this work, if the theory is expected to undergo grand unification in the ultraviolet. This is because the splitting $\delta$ is not a free parameter in the context of split SUSY. It is generated, in particular, by integrating out the bino and wino, leading to $\delta\simeq m_Z^2(s_W^2/M_1+c_W^2/M_2)$, so if the LZ event is from a higgsino its recoil energy measures the electroweak gaugino masses. Small splittings mean heavy gauginos: for same-sign masses $\delta=380$--$470$\,keV requires $M_2\geq14$--$17$\,PeV, or $M_1=M_2=18$--$22$\,PeV for a common mass, though opposite signs or phases allow cancellations and leave $M_2$ free. Such a spectrum affects the unification of couplings. Only the wino and the higgsino alter the gauge running coefficients $b_1-b_2$ at one loop, so $M_2$ fixes the scale at which $\alpha_1$ and $\alpha_2$ meet. At two loops, we find that $M_{\rm GUT} \propto M_2^{-1/4}$, with $M_{\rm GUT} \sim 10^{15}$\,GeV for $M_2 \sim 20$\,PeV.
The low unification scale is naively a problem because of proton decay in the context of a grand unified ultraviolet completion; additional threshold corrections would likely be needed to raise the masses of the proton-decay-generating fields. 
This danger was noted before at one loop for related spectra with a light
higgsino and heavy gauginos in Refs.~\cite{Cheung:2005ba,Jeong:2017rrs}.

\emph{Note added.}---LZ posted their analysis to the
arXiv~\cite{LZ:2026axp}  with a public data release while this Letter was
under review; the release gives the NEST parameters, the in-window
efficiency of Fig.~S2 ($50\%$ points at $5.4$ and $269.9$\,keV) and the
confidence intervals, but no further information on the higher-energy region.
Five analyses of the
event appeared on 2~September 2026. Ref.~\cite{Fan:2026higgsino} finds
$\delta\simeq300$--$380$\,keV under a SHM and notes that
an LMC-like tail pushes $\delta$ toward $500$\,keV.  They also  give the one-loop gauge running for this split SUSY spectrum.
Ref.~\cite{Freese:2026higgsino} finds that the fixed-coupling cross
section approaches LZ's published $\mathcal{O}_1$ interval at
$\delta\simeq350$\,keV; Ref.~\cite{Wu:2026higgsino} finds a best fit at
$324$\,keV with an Einasto profile, using Eddington inversion to obtain the velocity distribution;
Ref.~\cite{Su:2026inelastic} fits a generic vector-mediated state and
points to tungsten-based experiments for a future test; and Ref.~\cite{Lou:2026absorption} interprets the
event as absorption of a $247$\,MeV fermion and reports a significant
tension with their  recasting of KamLAND results. 

Six further interpretation papers appeared on 3~September 2026.
Ref.~\cite{Pospelov:2026SolarCapture} calculates constraints from neutrino observations of the Sun with IceCube, based on higgsinos (accelerated by the Sun's gravity to velocities $\simeq1400$\,km/s) being captured by scattering on iron and uranium, and subsequently annihilating. They find that the thermal higgsino can be excluded for $\delta\lesssim506$\,keV from tree-level up-scattering and
$\lesssim566$\,keV once the loop-induced spin-dependent channel is
included. The internal tension we have identified in the data is independent of and complementary to this constraint; however, lower DM masses that alleviate the tension with the higher-energy data may also weaken this solar limit, since IceCube's bound weakens for lower-energy neutrinos. We have also demonstrated that a high-velocity tail for the DM distribution can push the preferred splitting to higher values that will have less tension with the solar bounds, but achieving a preferred splitting over 566 keV at $m_\chi=1.1$\,TeV would require a larger effect from the high-velocity tail than any models we have tested in this work, and this might also be expected to enhance the tension with a lack of events in the high-energy control bin.  Note that Ref.~\cite{Pospelov:2026SolarCapture} also independently addresses the factor of four cross-section discrepancy that we discuss in the Appendix; their result agrees with ours.

Ref.~\cite{Yin:2026PQSUSY}
embeds $\delta\simeq0.28$--$0.35$\,MeV in high-scale SUSY and finds the
same low unification crossing that we do.
Ref.~\cite{DiMauro:2026KinematicEdge} obtains $\delta\simeq371$\,keV for
the thermal higgsino, matching our calculation with the Vietz form factor at $v_{\rm
esc}=544$\,km/s, and tests it with the gamma-ray line.
Refs.~\cite{Nomura:2026Z2Partner,Visinelli:2026PQInelastic,Yamashita:2026InelasticDP}
consider inelastic DM with mass in the range $380$--$600$\,GeV and mass splittings
$\delta\simeq316$--$350$\,keV, inside the lower-mass window that appears to alleviate the tension with our
sideband test.

\textit{Acknowledgments.}
NLR appreciated conversations with Ciaran O'Hare and Jayden Newstead.
We acknowledge the use of ChatGPT and Claude in helping prepare the numerical results and in cross checking the calculations provided in this work.
The research of NLR was supported by the Office of High Energy Physics of the U.S. Department of Energy under contract DE-AC02-05CH11231.
BRS is supported in part by the DOE award DESC0025293. 
BRS acknowledges support from the Alfred P. Sloan Foundation. TRS' work is supported by the Simons Foundation (Grant Number 929255, T.R.S) and by the U.S. Department of Energy, Office of Science, Office of High Energy Physics of U.S. Department of Energy under grant contract Number DE-SC0012567. The work of WLX
was supported by the Kavli Institute for Particle Astrophysics and Cosmology, and by the
U.S. Department of Energy under contract DE-AC02-76SF00515.

\appendix

\makeatletter
\par\addvspace{6pt plus 2pt}\noindent{\itshape Appendix A: Higgsinos away from the thermal mass}---\nobreak\ignorespaces
\makeatother
If the higgsino is all of the DM by non-thermal production, then $m_\chi$ may be treated as a free parameter. In Tab.~\ref{tab:mass} we display the relevant LZ-window quantities for benchmark masses $m_\chi = 500$ GeV--$100$ TeV at $v_{\rm esc} = 567$ km/s with the Helm response.
Over more than two decades in mass $\delta$ moves only from $359$ to $446$ keV, $\delta_{\rm max}$ saturates at $446$\,keV because $\mu_A \to m_{\rm Xe}$, and $v_{\rm min}$ at the surviving point stays between $739$ and $799$\,km/s at every mass. 
Note also that for $m_\chi \geq 500$ GeV the event energy alone would prefer $\delta = 248$--$309$\,keV ($E_R^* = 248$\,keV), always \emph{below} the $\delta$ that gives the right rate, so the event sits below the kinematic scale of the predicted spectrum. The predicted counts above the signal window at the $\delta$ that give one count, across the two responses and $v_{\rm esc} = 544$--$567$\,km/s, are $1.0$--$2.0$ at $500$\,GeV ($0.2$--$0.7$ in the high-energy sideband), $5.0$--$7.3$ at $800$\,GeV ($2.3$--$4.3$) and $9.1$--$12.3$ at $2.5$ TeV ($5.5$--$8.6$), so the sideband test impacts every SHM-tail higgsino above $\sim\!800$\,GeV. The $500$\,GeV benchmark is much less constrained by this, and we have not solved for the mass at which the transition occurs.  

\begin{table}[b]
\centering
\footnotesize
\setlength{\tabcolsep}{4.5pt}
\begin{tabular}{rrrrrr}
\toprule
$m_\chi$\,[GeV] & $\mu_A$\,[GeV] & $\delta_{\rm max}$\,[keV] & $\delta_1$\,[keV] & $E_R^*$ & $v_{\rm min}$\,[km/s] \\
\midrule
$500$ & $98.3$ & $359$ & $349$ & $281$ & $799$ \\
$1100$ & $110.1$ & $402$ & $377$ & $339$ & $784$ \\
$2500$ & $116.6$ & $426$ & $387$ & $369$ & $773$ \\
$5000$ & $119.4$ & $436$ & $389$ & $380$ & $765$ \\
$10^4$ & $120.8$ & $442$ & $387$ & $383$ & $759$ \\
$10^5$ & $122.1$ & $446$ & $371$ & $370$ & $739$ \\
\bottomrule
\end{tabular}
\caption{Higgsinos that are all of the DM at benchmark masses $m_\chi \neq 1.1$ TeV, on xenon, for LZ's $2.84$\,tonne-yr, $v_{\rm esc} = 544$ km/s, and the Helm response. Energies in keV; $E_R^*$ and $v_{\rm min}$ are evaluated at $\delta_1$, with $\delta_1$ the splitting that gives one expected count in the signal ROI.}
\label{tab:mass}
\end{table}

\par\addvspace{6pt plus 2pt}\noindent{\itshape Appendix B: The higgsino-nucleus scattering cross section}---\nobreak\ignorespaces
\makeatother
The cross section for higgsino-nucleus inelastic scattering, in the elastic limit where the mass splitting is small, is stated in Refs.~\cite{Nagata:2014wma,Bramante:2016rdh} to be:
\begin{align}
\sigma_\text{inelastic} = \frac{G_F^2}{8\pi} \mu^2 \left[N - (1 -4 s^2_W) Z \right]^2, 
\end{align}
where $\mu$ is the reduced mass of the nucleus-higgsino system, $Z$ is the atomic number and $N$ is the number of neutrons. However, this calculation is complicated by the Majorana nature of the higgsino states. 

To clarify the counting of degrees of freedom, it is helpful to note that in the limit of zero mass splitting, we may view the higgsino as a $SU(2)$ doublet of a charged Dirac fermion and a neutral Dirac fermion. The doublet has weak isospin $T^3=\pm 1/2$ and hypercharge $Y=1$, working in the convention where $Q=T^3 + Y/2$. We can compute the scattering of the neutral fermion off a nucleus at tree level; if we convert to the basis of Majorana fermions (which will be split in mass by the mass splitting), the tree-level scattering will be purely off-diagonal between the Majorana states, and so the off-diagonal scattering cross section must match up with the full scattering cross section for the neutral Dirac fermion. 

Ref.~\cite{Essig:2007az}, working in the same hypercharge convention, finds that the cross section for a chiral electroweakly-interacting DM particle is:
\begin{align}
\sigma = \frac{G_F^2}{2\pi} \mu^2 \left[ N - (1 -4 s^2_W) Z \right]^2 \bar{Y}^2, 
\end{align}
where $\bar{Y}=\frac{1}{2}(Y_L + Y_R)$ and $Y_L, Y_R$ are respectively the hypercharges of the left- and right-handed components of the multiplet. In the case of vector DM where $Y_L=Y_R$, $\bar{Y}=Y_L=Y_R$ and for the higgsino $\bar{Y}=1$. This result thus disagrees with Refs.~\cite{Nagata:2014wma,Bramante:2016rdh} by a factor of 4.

As there is no controversy about the relative coupling to protons and neutrons, we can focus on the case of neutrons alone to cross-check this factor of 4. In our convention, the neutron has hypercharge $Y=1/2$ and $T^3=-1/2$, and the neutral component of the higgsino multiplet also has $T^3=-1/2$.

The contribution to the vector coupling for an uncharged left-handed fermion field $f$ with isospin $T^3$ is $-g T^3/(2 c_W)$ (e.g.~\cite{ParticleDataGroup:2026mpi}); thus for the neutron this coupling is $g/(4 c_W)$. For the higgsino, there is an equal contribution from the right-handed component, so the coupling becomes $g/(2c_W)$. The neutron also has an axial vector coupling $-g/(4c_W)$; however, this term gives a spin-dependent cross section which will not be coherently enhanced. The spin-independent matrix element is then:
\begin{align} i \mathcal{M} & = 2 (-g/(4 c_W))^2 \bar{n}(p') \gamma^\mu n(p) \frac{-i g_{\mu \nu}}{(p' - p)^2 - m_Z^2}  \nonumber \\
& \bar{\chi}(k') \gamma^\nu \chi(k). \end{align} 
We can compute the unpolarized squared matrix element as usual and find:
\begin{align} |\mathcal{M}|^2  & = 4 (-g/(4 c_W))^4 \frac{1}{m_Z^4} (4 m_n m_\chi)^2 \nonumber \\
& = (G_F/\sqrt{2})^2 (4 m_n m_\chi)^2, \end{align}
where we use $G_F/\sqrt{2} = g^2/(8 m_W^2) = g^2/(8 c_W^2 m_Z^2)$. Consequently the cross section becomes:
\begin{align}\frac{d\sigma}{d\Omega} & = \frac{1}{64\pi^2 s} |\mathcal{M}|^2 \nonumber \\
\Rightarrow \sigma & = \frac{(4 m_n m_\chi)^2}{16\pi (m_\chi + m_n)^2}   \frac{G_F^2}{2}  =  \frac{1}{\pi} \mu_{n\chi}^2   \frac{G_F^2}{2}. \end{align} 
This is in agreement with the result of Ref.~\cite{Essig:2007az}.

\bibliographystyle{utphys}
\bibliography{refs}

@article{Bramante:2016rdh,
  author  = {Bramante, Joseph and Fox, Patrick J. and Kribs, Graham D. and Martin, Adam},
  title   = {Inelastic frontier: Discovering dark matter at high recoil energy},
  journal = {Phys. Rev. D},
  volume  = {94},
  pages   = {115026},
  year    = {2016},
  eprint  = {1608.02662},
  archivePrefix = {arXiv},
  primaryClass = {hep-ph},
  doi = {10.1103/PhysRevD.94.115026},
  number = {11}
}

@article{ParticleDataGroup:2026mpi,
    author = "Takahashi, F. and others",
    collaboration = "Particle Data Group",
    title = "{Review of Particle Physics}",
    doi = "10.1142/s0217751x26300115",
    journal = "Int. J. Mod. Phys. A",
    volume = "41",
    number = "22",
    pages = "2630011",
    year = "2026"
}

@article{Nagata:2014wma,
  author  = {Nagata, Natsumi and Shirai, Satoshi},
  title   = {Higgsino Dark Matter in High-Scale Supersymmetry},
  journal = {JHEP},
  volume  = {01},
  pages   = {029},
  year    = {2015},
  eprint  = {1410.4549},
  archivePrefix = {arXiv},
  primaryClass = {hep-ph},
  doi = {10.1007/JHEP01(2015)029}
}

@article{Essig:2007az,
  author  = {Essig, Rouven},
  title   = {Direct Detection of Non-Chiral Dark Matter},
  journal = {Phys. Rev. D},
  volume  = {78},
  pages   = {015004},
  year    = {2008},
  eprint  = {0710.1668},
  archivePrefix = {arXiv},
  primaryClass = {hep-ph},
  doi = {10.1103/PhysRevD.78.015004}
}

@article{Graham:2024hah,
  author  = {Graham, Peter W. and Ramani, Harikrishnan and Wong, Samuel S. Y.},
  title   = {Enhancing Direct Detection of Higgsino Dark Matter},
  journal = {Phys. Rev. D},
  volume  = {111},
  pages   = {055030},
  year    = {2025},
  eprint  = {2409.07768},
  archivePrefix = {arXiv},
  primaryClass = {hep-ph},
  doi = {10.1103/PhysRevD.111.055030},
  number = {5}
}

@article{Folsom:2025lly,
    author = "Folsom, Dylan and Blanco, Carlos and Lisanti, Mariangela and Necib, Lina and Vogelsberger, Mark and Hernquist, Lars",
    title = "{Dark Matter Velocity Distributions for Direct Detection: Astrophysical Uncertainties Are Smaller Than They Appear}",
    eprint = "2505.07924",
    archivePrefix = "arXiv",
    primaryClass = "hep-ph",
    doi = "10.1103/wmpq-mw4h",
    journal = "Phys. Rev. Lett.",
    volume = "135",
    number = "21",
    pages = "211004",
    year = "2025"
}

@article{RES-NOVA:2026,
  author        = {Alloni, D. and others},
  title         = {{Probing large mass-splitting inelastic Dark Matter with RES-NOVA}},
  eprint        = {2607.18378},
  archivePrefix = {arXiv},
  primaryClass  = {astro-ph.CO},
  year = {2026}
}

@article{Fan:2026higgsino,
  author  = {Fan, JiJi and Reece, Matthew},
  title   = {Higgsino Above the Sea of Fog},
  year    = {2026},
  eprint  = {2609.01504},
  archivePrefix = {arXiv},
  primaryClass = {hep-ph}
}

@article{Freese:2026higgsino,
  author        = {Freese, Katherine and Theodosopoulos, Dionysios P.},
  title         = {{Higgsino Dark Matter Interpretation of the LUX-ZEPLIN 248 keV Nuclear-Recoil Event}},
  eprint        = {2609.01583},
  archivePrefix = {arXiv},
  primaryClass  = {hep-ph},
  year = {2026}
}

@article{Wu:2026higgsino,
  author        = {Wu, Lei and Zhang, Yang and Zhu, Bin},
  title         = {{TeV Higgsino Dark Matter from LZ Nuclear Recoil to Fermi-LAT Gamma Rays}},
  eprint        = {2609.01590},
  archivePrefix = {arXiv},
  primaryClass  = {hep-ph},
  year = {2026}
}

@article{Su:2026inelastic,
  author        = {Su, Liangliang and Yang, Jin Min and Yang, Wen-Na},
  title         = {{Inelastic Dark Matter Signature at High Recoil Energy in LUX-ZEPLIN and CRESST}},
  eprint        = {2609.01475},
  archivePrefix = {arXiv},
  primaryClass  = {hep-ph},
  year = {2026}
}

@article{Lou:2026absorption,
  author        = {Lou, Yuanchao and Lu, Chih-Ting},
  title         = {{Fermionic Dark Matter Absorption and the High-Energy Event in LUX-ZEPLIN}},
  eprint        = {2609.01592},
  archivePrefix = {arXiv},
  primaryClass  = {hep-ph},
  year = {2026}
}

@article{Besla:2019xbx,
  author  = {Besla, Gurtina and Peter, Annika H. G. and Garavito-Camargo, Nicolas},
  title   = {The highest-speed local dark matter particles come from the Large Magellanic Cloud},
  journal = {JCAP},
  volume  = {11},
  pages   = {013},
  year    = {2019},
  eprint  = {1909.04140},
  archivePrefix = {arXiv},
  primaryClass = {astro-ph.GA},
  doi = {10.1088/1475-7516/2019/11/013}
}

@article{Smith-Orlik:2023kyu,
  author  = {Smith-Orlik, Adam and others},
  title   = {The impact of the Large Magellanic Cloud on dark matter direct detection signals},
  journal = {JCAP},
  volume  = {10},
  pages   = {070},
  year    = {2023},
  eprint  = {2302.04281},
  archivePrefix = {arXiv},
  primaryClass = {astro-ph.GA},
  doi = {10.1088/1475-7516/2023/10/070}
}

@article{Rodd:2024qsi,
    author = "Rodd, Nicholas L. and Safdi, Benjamin R. and Xu, Weishuang Linda",
    title = "{CTA and SWGO can discover Higgsino dark matter annihilation}",
    eprint = "2405.13104",
    archivePrefix = "arXiv",
    primaryClass = "hep-ph",
    doi = "10.1103/PhysRevD.110.043003",
    journal = "Phys. Rev. D",
    volume = "110",
    number = "4",
    pages = "043003",
    year = "2024"
}

@inproceedings{Wells:2003tf,
    author = "Wells, James D.",
    title = "{Implications of supersymmetry breaking with a little hierarchy between gauginos and scalars}",
    booktitle = "{11th International Conference on Supersymmetry and the Unification of Fundamental Interactions}",
    eprint = "hep-ph/0306127",
    archivePrefix = "arXiv",
    reportNumber = "MCTP-03-30",
    month = "6",
    year = "2003"
}

@article{Giudice:2004tc,
    author = "Giudice, G. F. and Romanino, A.",
    title = "{Split supersymmetry}",
    eprint = "hep-ph/0406088",
    archivePrefix = "arXiv",
    reportNumber = "CERN-PH-TH-2004-100",
    doi = "10.1016/j.nuclphysb.2004.08.001",
    journal = "Nucl. Phys. B",
    volume = "699",
    pages = "65--89",
    year = "2004",
    note = "[Erratum: Nucl.Phys.B 706, 487--487 (2005)]"
}

@article{Arkani-Hamed:2004ymt,
    author = "Arkani-Hamed, Nima and Dimopoulos, Savas",
    title = "{Supersymmetric unification without low energy supersymmetry and signatures for fine-tuning at the LHC}",
    eprint = "hep-th/0405159",
    archivePrefix = "arXiv",
    doi = "10.1088/1126-6708/2005/06/073",
    journal = "JHEP",
    volume = "06",
    pages = "073",
    year = "2005"
}

@article{Hall:2011jd,
    author = "Hall, Lawrence J. and Nomura, Yasunori",
    title = "{Spread Supersymmetry}",
    eprint = "1111.4519",
    archivePrefix = "arXiv",
    primaryClass = "hep-ph",
    reportNumber = "UCB-PTH-11-09",
    doi = "10.1007/JHEP01(2012)082",
    journal = "JHEP",
    volume = "01",
    pages = "082",
    year = "2012"
}

@article{Arvanitaki:2012ps,
    author = "Arvanitaki, Asimina and Craig, Nathaniel and Dimopoulos, Savas and Villadoro, Giovanni",
    title = "{Mini-Split}",
    eprint = "1210.0555",
    archivePrefix = "arXiv",
    primaryClass = "hep-ph",
    doi = "10.1007/JHEP02(2013)126",
    journal = "JHEP",
    volume = "02",
    pages = "126",
    year = "2013"
}

@article{Arkani-Hamed:2012fhg,
    author = "Arkani-Hamed, Nima and Gupta, Arpit and Kaplan, David E. and Weiner, Neal and Zorawski, Tom",
    title = "{Simply Unnatural Supersymmetry}",
    eprint = "1212.6971",
    archivePrefix = "arXiv",
    primaryClass = "hep-ph",
    month = "12",
    year = "2012"
}

@article{Bottaro:2022one,
    author = "Bottaro, Salvatore and Buttazzo, Dario and Costa, Marco and Franceschini, Roberto and Panci, Paolo and Redigolo, Diego and Vittorio, Ludovico",
    title = "{The last complex WIMPs standing}",
    eprint = "2205.04486",
    archivePrefix = "arXiv",
    primaryClass = "hep-ph",
    reportNumber = "CERN-TH-2022-080",
    doi = "10.1140/epjc/s10052-022-10918-5",
    journal = "Eur. Phys. J. C",
    volume = "82",
    number = "11",
    pages = "992",
    year = "2022"
}

@article{Abe:2025lci,
    author = "Abe, Shotaro and Inada, Tomohiro and Moulin, Emmanuel and Rodd, Nicholas L. and Safdi, Benjamin R. and Xu, Weishuang Linda",
    title = "{Discovering the Higgsino at CTAO-North within the Decade}",
    eprint = "2506.08084",
    archivePrefix = "arXiv",
    primaryClass = "hep-ph",
    doi = "10.1103/5qjj-l4b5",
    journal = "Phys. Rev. D",
    volume = "114",
    number = "2",
    pages = "023052",
    year = "2026"
}

@article{Fan:2013faa,
    author = "Fan, JiJi and Reece, Matthew",
    title = "{In Wino Veritas? Indirect Searches Shed Light on Neutralino Dark Matter}",
    eprint = "1307.4400",
    archivePrefix = "arXiv",
    primaryClass = "hep-ph",
    doi = "10.1007/JHEP10(2013)124",
    journal = "JHEP",
    volume = "10",
    pages = "124",
    year = "2013"
}

@article{Cohen:2013ama,
    author = "Cohen, Timothy and Lisanti, Mariangela and Pierce, Aaron and Slatyer, Tracy R.",
    title = "{Wino Dark Matter Under Siege}",
    eprint = "1307.4082",
    archivePrefix = "arXiv",
    primaryClass = "hep-ph",
    reportNumber = "MIT-CTP-4482, SLAC-PUB-15664, MCTP-13-19",
    doi = "10.1088/1475-7516/2013/10/061",
    journal = "JCAP",
    volume = "10",
    pages = "061",
    year = "2013"
}

@article{Safdi:2022xkm,
    author = "Safdi, Benjamin R.",
    title = "{TASI Lectures on the Particle Physics and Astrophysics of Dark Matter}",
    eprint = "2303.02169",
    archivePrefix = "arXiv",
    primaryClass = "hep-ph",
    doi = "10.22323/1.439.0009",
    journal = "PoS",
    volume = "TASI2022",
    pages = "009",
    year = "2024"
}

@article{Dessert:2022evk,
    author = "Dessert, Christopher and Foster, Joshua W. and Park, Yujin and Safdi, Benjamin R. and Xu, Weishuang Linda",
    title = "{Higgsino Dark Matter Confronts 14~Years of Fermi \ensuremath{\gamma}-Ray Data}",
    eprint = "2207.10090",
    archivePrefix = "arXiv",
    primaryClass = "hep-ph",
    reportNumber = "MIT-CTP/5454",
    doi = "10.1103/PhysRevLett.130.201001",
    journal = "Phys. Rev. Lett.",
    volume = "130",
    number = "20",
    pages = "201001",
    year = "2023"
}

@article{Rinchiuso:2020skh,
    author = "Rinchiuso, Lucia and Macias, Oscar and Moulin, Emmanuel and Rodd, Nicholas L. and Slatyer, Tracy R.",
    title = "{Prospects for detecting heavy WIMP dark matter with the Cherenkov Telescope Array: The Wino and Higgsino}",
    eprint = "2008.00692",
    archivePrefix = "arXiv",
    primaryClass = "astro-ph.HE",
    reportNumber = "MIT-CTP 5120, IRFU-20-13",
    doi = "10.1103/PhysRevD.103.023011",
    journal = "Phys. Rev. D",
    volume = "103",
    number = "2",
    pages = "023011",
    year = "2021"
}

@article{Chen:2019gtm,
    author = "Chen, Qing and Hill, Richard J.",
    title = "{Direct detection rate of heavy Higgsino-like and Wino-like dark matter}",
    eprint = "1912.07795",
    archivePrefix = "arXiv",
    primaryClass = "hep-ph",
    reportNumber = "FERMILAB-PUB-19-564-T",
    doi = "10.1016/j.physletb.2020.135364",
    journal = "Phys. Lett. B",
    volume = "804",
    pages = "135364",
    year = "2020"
}

@article{Co:2021ion,
    author = "Co, Raymond T. and Sheff, Benjamin and Wells, James D.",
    title = "{Race to find split Higgsino dark matter}",
    eprint = "2105.12142",
    archivePrefix = "arXiv",
    primaryClass = "hep-ph",
    reportNumber = "LCTP-21-11, UMN-TH-4016/21, FTPI-MINN-21/09",
    doi = "10.1103/PhysRevD.105.035012",
    journal = "Phys. Rev. D",
    volume = "105",
    number = "3",
    pages = "035012",
    year = "2022"
}

@article{Hill:2014yxa,
    author = "Hill, Richard J. and Solon, Mikhail P.",
    title = "{Standard Model anatomy of WIMP dark matter direct detection II: QCD analysis and hadronic matrix elements}",
    eprint = "1409.8290",
    archivePrefix = "arXiv",
    primaryClass = "hep-ph",
    reportNumber = "EFI-PREPRINT-14-25",
    doi = "10.1103/PhysRevD.91.043505",
    journal = "Phys. Rev. D",
    volume = "91",
    pages = "043505",
    year = "2015"
}

@article{Safdi:2025sfs,
    author = "Safdi, Benjamin R. and Xu, Weishuang Linda",
    title = "{Wino and Real Minimal Dark Matter Excluded by Fermi Gamma-Ray Observations}",
    eprint = "2507.15934",
    archivePrefix = "arXiv",
    primaryClass = "hep-ph",
    reportNumber = "CERN-TH-2025-137",
    month = "7",
    year = "2025"
}

@article{Cheung:2005ba,
  author  = {Cheung, Kingman and Chiang, Cheng-Wei and Song, Jeonghyeon},
  title   = {{A Minimal supersymmetric scenario with only mu at the weak scale}},
  journal = {JHEP},
  volume  = {04},
  pages   = {047},
  year    = {2006},
  eprint  = {hep-ph/0512192},
  archivePrefix = {arXiv},
  doi = {10.1088/1126-6708/2006/04/047},
  primaryClass = {hep-ph}
}

@article{Jeong:2017rrs,
  author  = {Jeong, Kwang Sik},
  title   = {{Light Higgsino for gauge coupling unification}},
  journal = {Phys. Lett. B},
  volume  = {769},
  pages   = {42},
  year    = {2017},
  eprint  = {1701.06947},
  archivePrefix = {arXiv},
  doi = {10.1016/j.physletb.2017.03.028},
  primaryClass = {hep-ph}
}

@article{Vietze:2014vsa,
  author        = {Vietze, L. and Klos, P. and Men{\'e}ndez, J. and Haxton, W.C. and Schwenk, A.},
  title         = {{Nuclear structure aspects of spin-independent WIMP scattering off xenon}},
  eprint        = {1412.6091},
  archivePrefix = {arXiv},
  primaryClass  = {nucl-th},
  doi           = {10.1103/PhysRevD.91.043520},
  journal       = {Phys. Rev. D},
  volume        = {91},
  number        = {4},
  pages         = {043520},
  year          = {2015}
}

@article{Anand:2013yka,
  author        = {Anand, Nikhil and Fitzpatrick, A. Liam and Haxton, W.C.},
  title         = {{Weakly interacting massive particle-nucleus elastic scattering response}},
  eprint        = {1308.6288},
  archivePrefix = {arXiv},
  primaryClass  = {hep-ph},
  doi           = {10.1103/PhysRevC.89.065501},
  journal       = {Phys. Rev. C},
  volume        = {89},
  number        = {6},
  pages         = {065501},
  year          = {2014}
}

@article{Baxter:2021pqo,
  author        = {Baxter, D. and Bloch, I. M. and Bodnia, E. and others},
  title         = {{Recommended conventions for reporting results from direct dark matter searches}},
  eprint        = {2105.00599},
  archivePrefix = {arXiv},
  primaryClass  = {hep-ex},
  doi           = {10.1140/epjc/s10052-021-09655-y},
  journal       = {Eur. Phys. J. C},
  volume        = {81},
  number        = {10},
  pages         = {907},
  year          = {2021}
}

@article{Lewin:1995rx,
    author = "Lewin, J. D. and Smith, P. F.",
    title = "{Review of mathematics, numerical factors, and corrections for dark matter experiments based on elastic nuclear recoil}",
    doi = "10.1016/S0927-6505(96)00047-3",
    journal = "Astropart. Phys.",
    volume = "6",
    pages = "87--112",
    year = "1996"
}

@article{Feldman:1997qc,
    author = "Feldman, Gary J. and Cousins, Robert D.",
    title = "{A Unified approach to the classical statistical analysis of small signals}",
    eprint = "physics/9711021",
    archivePrefix = "arXiv",
    reportNumber = "HUTP-97-A096",
    doi = "10.1103/PhysRevD.57.3873",
    journal = "Phys. Rev. D",
    volume = "57",
    pages = "3873--3889",
    year = "1998"
}

@article{Pospelov:2026SolarCapture,
  author        = {Pospelov, Maxim and Ramani, Harikrishnan},
  title         = {{Strong Constraints on Higgsino Dark Matter from Solar Capture}},
  eprint        = {2609.02775},
  archivePrefix = {arXiv},
  primaryClass  = {hep-ph},
  year          = {2026}
}

@article{Yin:2026PQSUSY,
  author        = {Yin, Wen},
  title         = {{A PQ-Symmetric High-Scale SUSY Interpretation of the LZ High-Energy Recoil}},
  eprint        = {2609.01892},
  archivePrefix = {arXiv},
  primaryClass  = {hep-ph},
  year          = {2026}
}

@article{DiMauro:2026KinematicEdge,
  author        = {Di Mauro, Mattia},
  title         = {{Dark Matter at the Kinematic Edge: Interpreting the 248 keV LZ Nuclear-Recoil Candidate}},
  eprint        = {2609.02608},
  archivePrefix = {arXiv},
  primaryClass  = {hep-ph},
  year          = {2026}
}

@article{Nomura:2026Z2Partner,
  author        = {Nomura, Yasunori},
  title         = {{Dark Matter as the $Z_2$ Partner of the Standard Model Higgs Boson}},
  eprint        = {2609.02505},
  archivePrefix = {arXiv},
  primaryClass  = {hep-ph},
  year          = {2026}
}

@article{Visinelli:2026PQInelastic,
  author        = {Visinelli, Luca},
  title         = {{A Peccei--Quinn Origin for Inelastic Electroweak Dark Matter after LUX-ZEPLIN}},
  eprint        = {2609.02807},
  archivePrefix = {arXiv},
  primaryClass  = {hep-ph},
  year          = {2026}
}

@article{Yamashita:2026InelasticDP,
  author        = {Yamashita, Kimiko},
  title         = {{Inelastic Dark Photon Dark Matter for the LUX-ZEPLIN High-Recoil Event and the Galactic Halo Gamma-Ray Excess}},
  eprint        = {2609.02868},
  archivePrefix = {arXiv},
  primaryClass  = {hep-ph},
  year          = {2026}
}

@article{deSalas:2020hbh,
    author = "de Salas, Pablo F. and Widmark, Axel",
    title = "{Dark matter local density determination: recent observations and future prospects}",
    eprint = "2012.11477",
    archivePrefix = "arXiv",
    primaryClass = "astro-ph.GA",
    doi = "10.1088/1361-6633/ac24e7",
    journal = "Rept. Prog. Phys.",
    volume = "84",
    number = "10",
    pages = "104901",
    year = "2021"
}

@article{Monari:2018ckf,
    author = "Monari, G. and Famaey, B. and Carrillo, I. and Piffl, T. and Steinmetz, M. and Wyse, R. F. G. and Anders, F. and Chiappini, C. and Jan{\ss}en, K.",
    title = "{The escape speed curve of the Galaxy obtained from Gaia DR2 implies a heavy Milky Way}",
    eprint = "1807.04565",
    archivePrefix = "arXiv",
    primaryClass = "astro-ph.GA",
    doi = "10.1051/0004-6361/201833748",
    journal = "Astron. Astrophys.",
    volume = "616",
    pages = "L9",
    year = "2018"
}

@article{Necib:2021vxr,
    author = "Necib, Lina and Lin, Tongyan",
    title = "{Substructure at High Speed. II. The Local Escape Velocity and Milky Way Mass with Gaia eDR3}",
    eprint = "2102.02211",
    archivePrefix = "arXiv",
    primaryClass = "astro-ph.GA",
    doi = "10.3847/1538-4357/ac4244",
    journal = "Astrophys. J.",
    volume = "926",
    number = "2",
    pages = "189",
    year = "2022"
}

@article{Necib:2021yhq,
    author = "Necib, Lina and Lin, Tongyan",
    title = "{Substructure at High Speed. I. Inferring the Escape Velocity in the Presence of Kinematic Substructure}",
    eprint = "2102.01704",
    archivePrefix = "arXiv",
    primaryClass = "astro-ph.GA",
    doi = "10.3847/1538-4357/ac4243",
    journal = "Astrophys. J.",
    volume = "926",
    number = "2",
    pages = "188",
    year = "2022"
}

@article{Folsom:2026dqs,
    author = "Folsom, Dylan and Blanco, Carlos and Lisanti, Mariangela and Vogelsberger, Mark",
    title = "{Ubiquitous Corotation of Dark Matter Halos: Implications for Direct Detection}",
    eprint = "2608.00161",
    archivePrefix = "arXiv",
    primaryClass = "hep-ph",
    month = "7",
    year = "2026"
}

@article{LZ:2026axp,
    author = "Akerib, D. S. and others",
    collaboration = "LZ",
    title = "{Search for dark matter particle interactions in an extended nuclear recoil energy window with the LUX-ZEPLIN (LZ) experiment}",
    eprint = "2609.02823",
    archivePrefix = "arXiv",
    primaryClass = "hep-ex",
    month = "9",
    year = "2026"
}

@article{OHare:2021utq,
    author = "O'Hare, Ciaran A. J.",
    title = "{New Definition of the Neutrino Floor for Direct Dark Matter Searches}",
    eprint = "2109.03116",
    archivePrefix = "arXiv",
    primaryClass = "hep-ph",
    doi = "10.1103/PhysRevLett.127.251802",
    journal = "Phys. Rev. Lett.",
    volume = "127",
    number = "25",
    pages = "251802",
    year = "2021"
}

\end{document}